\documentclass{PoS}
\def\grs{GRS $1915$+$105$}
\def\ergcms{erg cm$^{-2}$ s$^{-1}$ }
\def\integral{{\it{INTEGRAL}}}
\def\rxte{{\it{RXTE}}}
\usepackage{graphicx}
\usepackage{epsfig,epsf}

\title{An INTEGRAL monitoring of GRS 1915+105 using simultaneous space and ground based instruments}
\ShortTitle{A multi-wavelength monitoring of GRS 1915+105}
\author{\speaker{J\'er\^ome Rodriguez}$^1$, Guy Pooley$^2$, Diana Hannikainen$^3$, Harry J. Lehto$^4$, Tomaso Belloni$^5$, Marion Cadolle-Bel$^1$, St\'ephane Corbel$^{6,1}$\\
$^1$CEA Saclay, France\\ 
$^2$Cavendish Laboratory, University of Cambridge, UK\\
$^3$Observatory, University of Helsinki, Finland\\
$^4$Tuorla Observatory, Turku Finland \&  Nordita, Denmark\\
$^5$Oss. Astronomico di Brera, Merate, Italy\\
$^6$Universit\'e Paris 7, France\\
        E-mail: \email{jrodriguez@cea.fr}}

\abstract{We report the results of 3 observations of GRS 1915+105 during which the 
source is found to show the X-ray dips/spike sequences (cycles). These observations 
 were performed simultaneously with \integral, 
\rxte, the Ryle and Nan\c{c}ay radio telescopes. They  show  
the so-called $\nu$, $\lambda$ and $\beta$ classes of variability during which  
a high level of correlated X-ray and  radio variability is observed.
We study the connection between the accretion processes seen in the X-ray 
behaviour, and the ejections seen in radio. 
By observing ejection during class $\lambda$, we generalise the fact that 
the discrete ejections in GRS 1915+105 occur after the cycles seen at X-ray 
energies, and identify the most likely trigger of the ejection through a 
spectral approach to our \integral\ data.  We show that each ejection is 
very probably the result of the ejection of a Comptonising medium 
responsible for the hard X-ray emission seen above 15 keV with \integral.
We study the rapid variability of the source, and observe the presence of Low
Frequency Quasi Periodic Oscillations during the X-ray dips.  The ubiquity of the 
former behaviour, and the following ejection may 
suggest a link between the QPO and the mechanism responsible for the 
ejection. }

\FullConference{VI Microquasar Workshop: Microquasars and Beyond\\
		 September 18-22, 2006\\
		 Como, Italy}
\begin{document}

\section{Introduction}
GRS~1915+105 is probably the most spectacular high-energy source of our Galaxy.
An extensive review on it can be found in Fender \& Belloni (2004).
To summarize, \grs\ is a microquasar hosting a black hole (BH) of  
14.0 $\pm$ 4.4 M$_{\odot}$ (Harlaftis \& Greiner 2004),
it is one of the brightest X-ray sources in the sky and it is a source 
of superluminal ejection (Mirabel \& Rodriguez 1994), with true velocity 
of the jets $\geq 0.9c$.
 The source is also known to show a compact jet during its periods of 
low steady levels of emission (e.g. Fuchs et al.\ 2003). 
Multi-wavelength coverage from radio to X-ray  has shown a
clear but complex association between the soft X-rays and radio/IR behaviour.
Of particular relevance is the existence  of radio QPO in the range 
20--40 min  associated with the X-ray variations on the same time scale 
(e.g. Mirabel et al. 1998). These
so called ``30-minute cycles'' were interpreted as being due 
to small ejections of material from the
system, and were found to correlate with the disc instability, as
observed in the X-ray band. \\
\indent Extensive  observations at X-ray energies with  {\it RXTE} 
have allowed   
Belloni et al.\ (2000) to  classify all the observations into
12 separate classes (labeled with greek letters), which could be 
interpreted as transitions between
three basic states (A-B-C): a hard state and two softer states.  These
spectral changes are, in most classes, interpreted as reflecting 
the rapid disappearance of the inner portions of an accretion disc, followed
 by a slower refilling of the emptied region (Belloni et al.\ 1997).  \\
\indent In the X-ray timing domain, \grs\ also shows interesting features, 
such as the presence of Low or High Frequency Quasi-Periodic 
Oscillations (LFQPO, HFQPO) whose presence is, as observed in other 
microquasars, tightly linked to the X-ray behaviour.
LFQPOs with variable frequency during classes 
showing cycles have been reported. Correlations between the frequency and some of the 
spectral parameters have been pointed out (e.g. Markwardt et al.\ 1999, 
Rodriguez et al.\ 2002a,b). 
\grs\ is also 
one of the first BH systems in which the presence of HFQPOs has been 
observed, first at 
$\sim65-69$ Hz, and up to $\sim 170$~Hz (Morgan, Remillard \& Greiner, 1997; 
Belloni et al.\ 2006).\\
\indent The link between the accretion and ejection processes is, however, 
far from  understood and different kinds of model are proposed to explain all 
observational facts also including the X-ray low (0.1-10 Hz) frequency QPOs. 
 The monitoring campaign we have performed since late 
2002 with \integral\ and other instruments (mainly \rxte\ and the Ryle 
Telescope, but also Spitzer, Nan\c{c}ay, GMRT, Suzaku; see Ueda et al. these 
proceedings for the report of the latter) has the aim to try to understand the 
physics of the accretion-ejection phenomena, including,  for the first time, 
the behaviour of
the source seen above 20 keV up to few hundred keV. 
We report here the results obtained during observations performed 
during AO2 and AO3 
showing sequences of  X-ray hard dips/soft spikes (the cycles), 
followed by radio flares. The X-ray spectral results of these observations 
can be found in Rodriguez et al. (2006), while the full details 
of these analysis and extended discussions will be presented in a forthcoming 
paper (Rodriguez et al.in prep.).

\section{Description of the 3 observations}
\begin{figure}[!t]
\centering
\epsfig{file=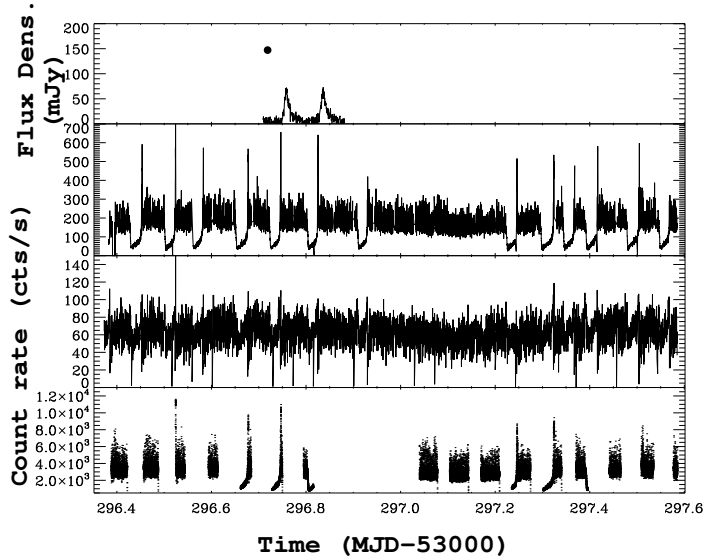,width=10cm}
\caption{Multiwavelength light curves of \grs\ on October 17-18 2004 showing two intervals 
of class $\nu$ variability. From top to bottom: Ryle (line) at 15 GHz, and Nan\c{c}ay (point) at 2.7 GHz, 
 JEM-X 3-13 keV, ISGRI 18-100 keV, \rxte\ 2-60 keV.}
\label{fig:Oct04}
\end{figure}

\begin{figure}[!t]
\centering
\epsfig{file=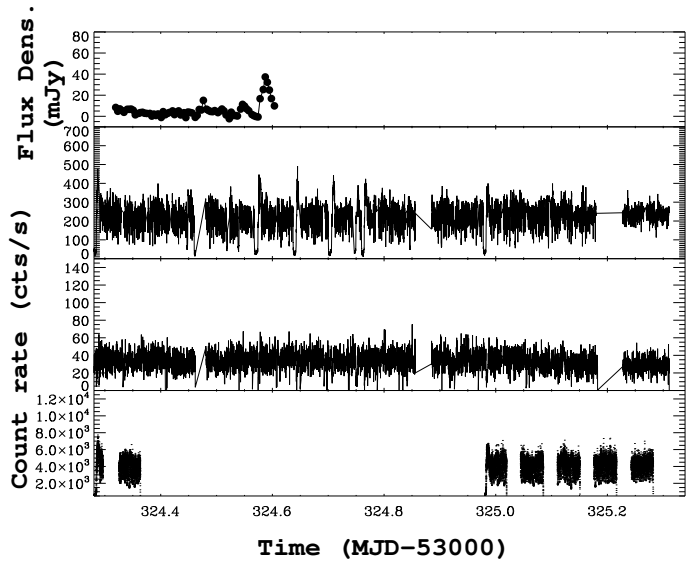,width=10cm}
\caption{Multiwavelength light curves of \grs\ on November 2004 during class $\lambda$ variability.
 The panels are the same as in Fig. 1 with the exception that no Nan\c{c}ay data are available.}
\label{fig:Nov04}
\end{figure}

\begin{figure}
\centering
\epsfig{file=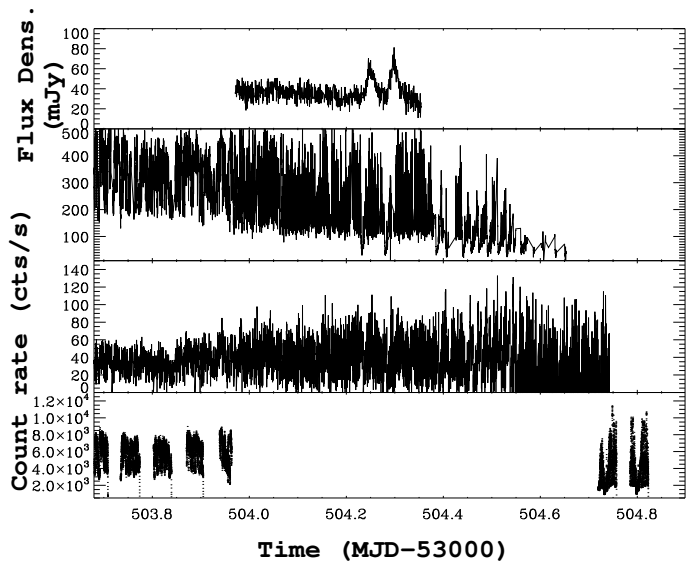,width=10cm}
\caption{Multiwavelength light curves of \grs\ on May 2005 with the appearance of  class $\beta$
variability half way through the observation. The panels 
are the same as in Fig. 2.}
\label{fig:May05}
\end{figure}

We focus in this paper on 3 observations respectively taken on 
October 17-18 2004 (Obs. 1), November 
15-16 2004 (Obs. 2), and May 13-14 2005 (Obs. 3). The multiwavelength 
light curves are shown in  Fig. \ref{fig:Oct04} to \ref{fig:May05}.
The JEM-X light curves show in all cases the occurrences of soft X-ray dips of different duration,
followed by a short spike marking the return to a high degree of soft X-ray emission and 
variability (hereafter cycle). In all the following we focus on the moments of 
cycle activity and more specifically on the intervals during which we have simultaneous 
data at radio wavelengths, and/or 
\rxte\ data for the timing analysis. This is particularly relevant for Obs.1 and 3 during which 
transitions between different classes are seen. We, therefore, avoid including 
in our analysis classes for which the radio behaviour is not known.
These observations are  classified following Belloni et al. (2000) as $\nu$, 
$\lambda$, and $\beta$ type cycle activity for Obs. 1, 2 and 3, respectively. 
In all three cases we observe at least one radio flare, which is indicative of a small  ejection 
of material (Mirabel et al. 1998), after a cycle. The  high flux shown by 
Nan\c{c}ay at 2.7 GHz during the class $\nu$ observation suggests that the radio flares follow  
systematically 
each cycle, as reported by Klein-Wolt et al. (2002). The presence of radio oscillations is also 
known in class $\beta$,
while no radio/X-ray connection had ever been observed during a class $\lambda$.\\
 
\section{X-ray Spectral Analysis}
\subsection{Selection of Good Time Intervals}
In order to study the spectral evolution of \grs\ through the cycle and the 
possible origin of the radio ejection, we divided the cycles into different intervals 
from which JEM-X and ISGRI spectra were accumulated. Each cycle was divided in, at least, 
3 intervals based on the soft X-ray count rate and the 3-13/18-100 keV hardness ratio. 
The intervals 
are defined as follow:
\begin{figure}
\centering
\epsfig{file=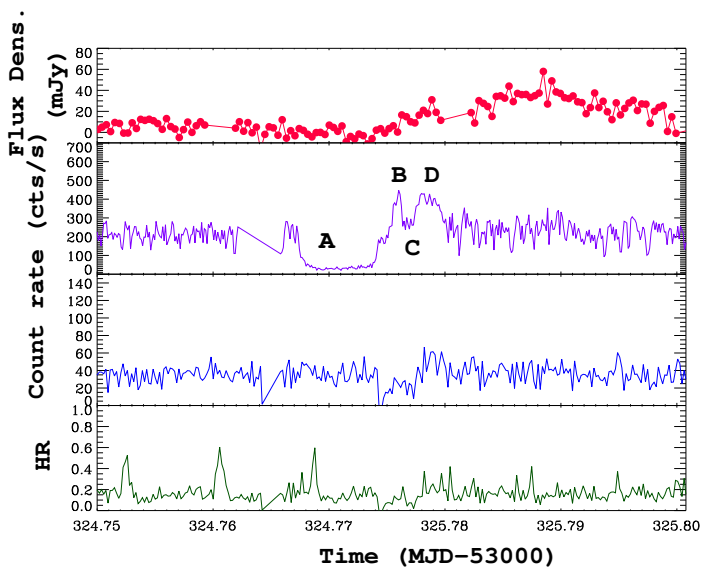,width=10cm}
\caption{A section on the X-ray cycle followed by the radio flare from class $\lambda$. 
The 4 intervals from which 
the spectra were accumulated (see text) are indicated. From top to bottom, the panels 
respectively represent the 
Ryle light curve, the 3-13 keV JEM-X light curve, the 18-100 keV ISGRI light curve, and the 18-100/3-13 keV hardness 
ratio.}
\label{fig:zoom}
\end{figure}

\begin{figure}[!t]
\centering
\epsfig{file=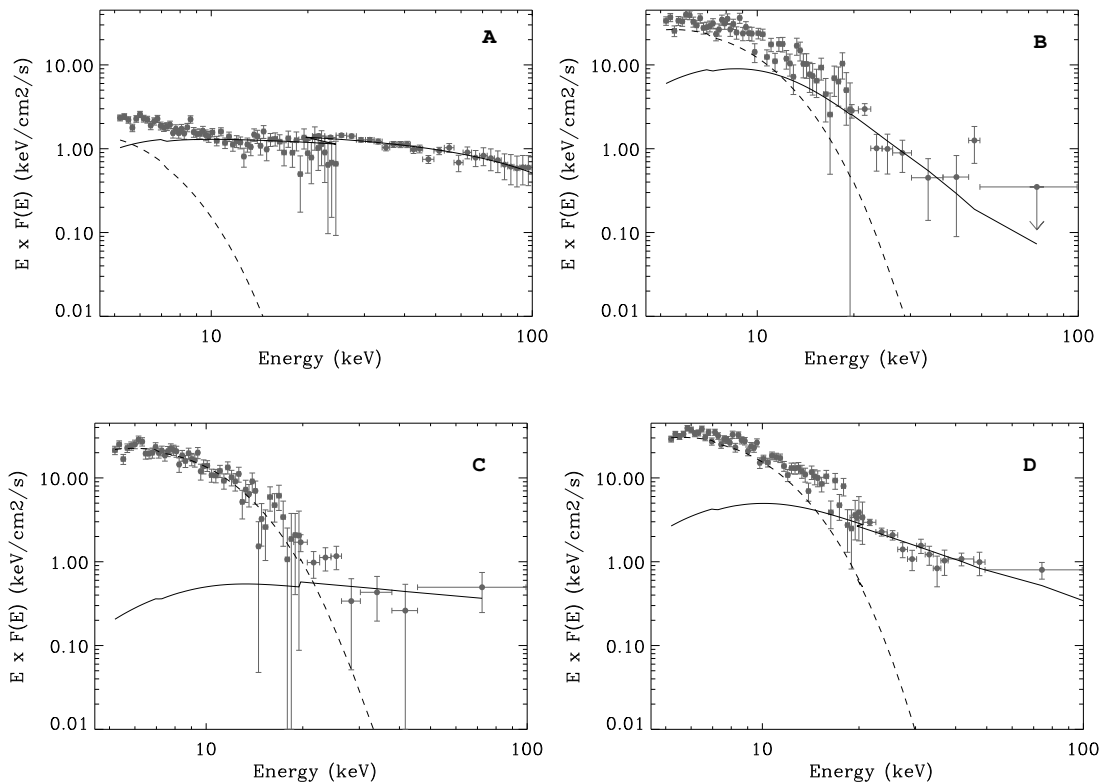,width=15cm}
\caption{The 4 spectra and best models superimposed from class $\lambda$. The individual components 
are also shown.
The disc component is represented by the 
 dashed line, while the Comptonised component is represented by the solid line. The evolution 
through the dips is quite obvious, with in particular a spectacular drop in the Comptonised 
component between B and C.}
\label{fig:spec}
\end{figure}

In each cycle four distinct intervals can be distinguished. The soft X-ray dip, having a hard
spectrum (interval A), a short precursor spike (interval B), a (short) following dip 
(interval C) and the main spike  (interval D), the last 3 having soft spectra. 
In class $\lambda$ and $\beta$ the observation 
of radio oscillations (Klein-Wolt et al. 2002), i.e. the occurrence of radio flares after each cycle, 
allow us to further accumulate together all cycles before the fitting to increase the 
statistics. In Class $\lambda$, occurrences
of a pairs of cycles at some moments can happen, while at other intervals the cycles are  isolated. Since no 
X-ray/radio connection had ever been observed in this class before, we extracted spectra from the 
unique cycle that 
is  followed by an ejection. The same sequence of A, B, C, D interval can also be identified 
(Fig. 4). 
Note the A, B, C, D labels for the different intervals are not related to the spectral states (A, B
and C) identified by Belloni et al. (2000). 
In order to be able to compare the different spectra, we fitted them all with 
the same model 
consisting of a thermal component ({\tt{ezdisk}}) and a Comptonised one ({\tt{comptt}}). 
Fig. \ref{fig:spec} shows,
as an example, the sequence of spectra from class $\lambda$.

\subsection{Results}
In all cases the evolution of the source through the cycles is more prononced in the soft
 X-rays (Fig. \ref{fig:Oct04}
to \ref{fig:May05}). At first sight we could think that the variations are then caused by changes 
in the accretion disc rather than in the Comptonised component. When fitting the different 
spectra with physical 
models, however, it seems to be the contrary. In class $\nu$ and $\lambda$ the evolution through  A to 
D translates into an apparent approach of the accretion disc to the black hole, with an increasing temperature. Interestingly between 
B and C, the disc parameters are compatible with being constant within the errors. We calculated
the 3-50 keV unabsorbed fluxes of the two spectral components in the 4 intervals.  While the 
flux from the 
disk increases through the whole sequence, that of the Comptonised component is not that regular. It first increases 
from $\nu_A$ to $\nu_B$ before decreasing by a factor of 2.5 in $\nu_C$ (reaching $1.0\times 10^{-8}$ \ergcms), 
and slowly recovers in $\nu_D$ ($1.5\times 10^{-8}$ \ergcms). It follows similar evolution in $\lambda$ 
(this is exemplified by Fig. \ref{fig:spec}), with in particular a reduction by a factor of $\sim 11$ between 
$\lambda_B$ and $\lambda_C$ (from $2.2\times 10^{-8}$ \ergcms to $0.19\times 10^{-8}$ \ergcms, when leaving 
all spectral parameters free to vary), with a lower limit of 2.7 (if the disc temperature is frozen to 
the same value as in $\lambda_B$). Again after $\lambda_C$
the Comptonised component slightly recovers ($1.32\times 10^{-8}$ \ergcms).\\
\indent The spectral behaviour of the source during class $\beta$ has been extensively 
studied in the past. It is interesting
to note that, as for the previous two classes, the disc temperature slowly increases during 
the dip, even after the 
spike (Swank et al. 1998), while the inner radius remains small, indicating the disc is 
close to the compact object.
The transition, after the spike, to state A
(Belloni et al. 2000) the very soft spectral state of \grs, suggest that the spectacular
 changes occurring at the spike are related to the Comptonised component.

\section{Timing analysis: Low Frequency QPOs}
\begin{figure}
\centering
\epsfig{file=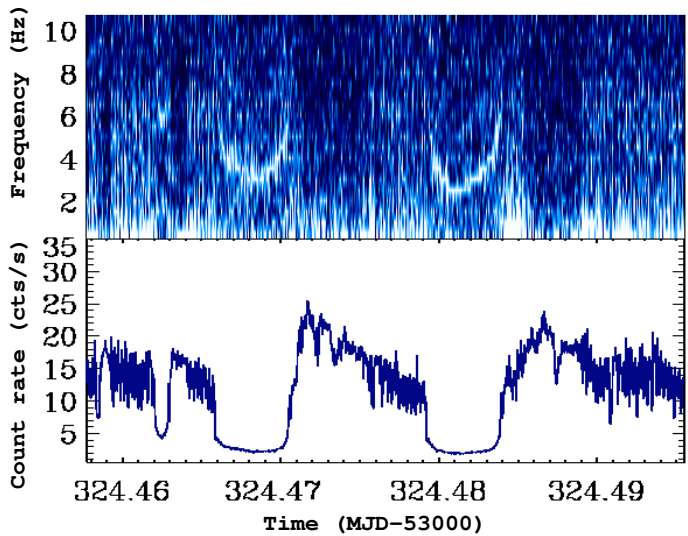,width=10cm}
\caption{Dynamical power spectrum (top) and \rxte/PCA light curve (bottom) on the unique 2 
cycles from class $\lambda$ covered by \rxte.}
\label{fig:dynpo2}
\end{figure}

Although most of the cycles seen with \rxte\ were not simultaneous with the 
cycles followed by radio ejections, we produced dynamical power spectra of the cycles 
with the view to study whether or not the soft X-ray dips were associated with 
a LFQPO of variable frequency,  as has been seen in class $\beta$, and $\alpha$, $\nu$
and $\theta$ (Rodriguez et al.2002a,b; Vadawale 2003). The dip preceding the radio ejection during class 
$\lambda$ was unfortunately 
not covered by \rxte. Assuming the same pattern repeats during all dips, we extracted 
a dynamical power spectrum from the unique sequence of dips covered by \rxte.  This is 
shown in Fig. \ref{fig:dynpo2} together with the associated light curve. In the other classes 
i.e. $\lambda$ and $\beta$, the presence of LFQPOs with variable frequencies during the dips
are confirmed by our analysis (not shown here).\\

\indent As in the other classes showing soft X-ray dips, a strong QPO with variable frequency 
appears during the dip (Fig. \ref{fig:dynpo2}). Here, however, the frequency seems less tightly
correlated to the X-ray count rate, since while during the dips the count rate remains rather 
constant, the frequency of the QPO is less stable.

\section{Discussion and conclusion}
We have presented observations of \grs\ made in simultaneity with \integral, \rxte, the Ryle 
telescope
and in one occasion the Nan\c{c}ay telescope. These observations belong to classes 
$\nu, \lambda$, and $\beta$, which have the characteristic of showing sequences of soft 
X-ray dips followed 
by high level of X-ray emission with high variability, sequences we referred to as cycles. 
While in some of these classes ejections (observed in the radio domain)  had been observed 
to follow to recovery to a high 
level of X-ray emission (Mirabel et al. 1998), the observation of a radio flare 
during the class $\lambda$ observation is, to our knowledge,  the first ever reported. This allows 
us to generalise the fact that there is always an ejection following a cycle. The duration of the 
dip during class $\lambda$ is 500s, 
while it is as low as 330s during class $\beta$. This duration is in agreement with the results
of Klein-Wolt et al. (2002) who observed that for an ejection to occur a longer than 100s 
hard state must occur.\\
\indent Detailed inspection of the cycles shows that the sequences are not as simple as 
it seems at 
first sight, but that in all cases a precursor spike occurs prior to the recovery to the high 
X-ray level. Our spectral analysis of the different portions of the cycles shows that the same 
evolution roughly occurs during the cycles from different classes. During the X-ray dip the source 
is in its hard state (or so-called C state for \grs\ Belloni et al. 2000). The precursor spikes 
correspond to a simultaneous softening of the source spectrum and an increase in the flux of the 
two main spectral components, the disc and the corona. The short dip immediately following the 
precursor corresponds to the  disappearance of the corona in the two cases we analysed in detail.
 The 
same had been observed in the case of class $\beta$ (Chaty 1998).
Given the observation of ejected material soon after, we conclude that:\\
1) {\bf{The ejected material is the coronal material}}\\
2) {\bf{The true moment of the ejection is the precursor spike}}\\
It has to be noted that Chaty (1998), Rodriguez et al. (2002b),Vadawale (2003) had come to similar 
conclusions during 
other classes. Another point compatible with this scenario is that if we estimate the lag between 
the radio and the X-ray, we obtain 0.28h and 0.26h during class $\nu$, 0.31h in class $\lambda$, 
and 
0.29 and 0.34h during class $\beta$, therefore the delay is  comparable in all classes. 
Interestingly 
the same conclusion had been drawn in another microquasar XTE J1550$-$564. During its 2000 
outburst (Rodriguez et al. 2003) have shown that the discrete ejection following the maximum of the X-ray 
emission was compatible with the ejection of the coronal medium.\\
\indent Although the physical properties of the dips are not exactly the same, some other 
similarities raise interesting questions. The timing analysis of the classes showing cycles 
all show the presence of a transient LFQPO appearing during the dip, and with a variable frequency 
which may be somehow correlated to the X-ray flux (Swank et al. 1998; Rodriguez et al. 2002a,b; Vadawale 2003). 
The presence of LFQPO during hard states is quite common, and may be a signature of the  
physics occurring during the accretion and the ejection of matter, since a steady compact 
jet is usually associated with this state. In the case of the cycles, a 
 discrete ejection seems to take place at the end of each cycle. The presence of 
LFQPO during the dip may, again, indicate that the QPO phenomenom and the accretion-ejection 
physics are linked. \\
\indent Recently Tagger et al. (2004) have proposed a ``magnetic flood'' scenario in trying to explain 
the occurrence of dips and ejections, based on a magnetic instability, the Accretion Ejection 
Instability (AEI, Tagger \& Pellat 1999). This instability has the effect of transporting angular 
momentum and energy from the inner region of the disc, and emitting them perpendicularly 
and directly into the corona.
A strong observational signature of this AEI is the presence of LFQPOs. 
In this framework the X-ray 
dip would correspond to the appearance of the AEI (equivalent to a transition to a hard state 
with appearance of LFQPO), and the ejection could be due to a reconnection event in the inner 
region of the disc (Tagger et al. 2004). An effect of the latter would be to blow the corona and would 
end observationally as an ejection. Although none of our results brings proof of such a 
scenario, and 
that other models exist, this scenario is compatible with our results. In particular with the 
generalisation of the ejection during all classes with dips, the relative similarities between 
the dips of all classes, both at radio and X-ray energies, and the constant presence of LFQPO 
during the dips, we feel that this model is a very promising one.

\section*{Acknowledgments}
These results are presented on behalf of a much larger collaboration whose members the
authors deeply thank.
JR is extremely grateful to J. Chenevez, and C.-A. Oxborrow for their precious help with the 
JEM-X data reduction, and M. Tagger for a careful reading of an early version of this paper. 
JR acknowledges E. Kuulkers and E. Smith and more generally the \integral\ 
and \rxte\ planning teams for their great efforts to have both satellites observing GRS 1915+105 
and IGR J19140+0951 at the same time.


\begin{thebibliography}{99}
 \bibitem{belloni97} 
Belloni, T., Mendez, M., King, A. R., et al. 1997, ApJ, 488, 109
\bibitem{belloni00} 
Belloni, T., Klein-Wolt, M., Mendez, et al. 2000, A\&A, 355, 271
\bibitem{belloni06} 
Belloni, T., Soleri, P., Casella, P., M\'endez, M., Migliari, S. 2006, MNRAS 369, 305.
\bibitem{chaty98} Chaty, S. 1998, Th\`ese de Doctorat ``Etude multi-longueur d'onde du microquasar GRS 1915+105 et de sources binaires.de haute \'energie de la Galaxie''
\bibitem{fender04} 
Fender, R.P. \& Belloni, T. 2004, ARA\&A, 42, 317
\bibitem{fuchs03} 
Fuchs, Y., Rodriguez, J., Mirabel, F., et al. 2003, A\&A, 409, L35.
\bibitem{harl04} 
Harlaftis, E.T \& Greiner, J., 2004, A\&A, 414, 13.
\bibitem{klein02} 
Klein-Wolt, M., Fender, R. P.; Pooley, G. G., et al. 2002, MNRAS, 331, 745
\bibitem{markwardt99}
Markwardt, C.B., Swank, J. H., Taam, R. E. 1999, ApJ, 513, 37.
\bibitem{m&r94} 
Mirabel, I.F. \& Rodr\'\i guez, L.F. 1994, Nature, 371, 46 
\bibitem{mirabel98} 
Mirabel, I.F., Dhawan, V., Chaty, S., et al. 1998, A\&A, 330, L9
\bibitem{morgan} 
Morgan, E.H., Remillard, R.A \& Greiner, J. 1997, ApJ, 482, 993 
\bibitem{rod02a} 
Rodriguez, J., Durouchoux, P., Tagger, M., et al., 2002a, A\&A, 386, 271
\bibitem{rod02b} 
Rodriguez, J., Varni\`ere, P., Tagger, M., Durouchoux, P. 2002b, A\&A, 387, 487
\bibitem{rod03} 
Rodriguez, J., Corbel, S., \& Tomsick, J.A.,  2003, ApJ, 595, 1032
\bibitem{rodri06} 
Rodriguez, J., Pooley, G., Hannikainen, D.C., Lehto, H.J.  2006, proceedings 
of the 6th INTEGRAL Workshop, ESA SP622.
\bibitem{swank98}
Swank, J.H., Chen, X., Markwardt, C., Taam, E. 1998 proceedings of the conference "Accretion Processes in Astrophysics: Some Like it Hot", eds. S. Holt and T. Kallman
\bibitem{tagger99} 
Tagger, M. \& Pellat, R. 1999, A\&A, 349, 1003
\bibitem{tagger04} 
Tagger, M., Varni\`ere, P., Rodriguez, J. \& Pellat, R. 2004, ApJ, 607,  410
\bibitem{vadawale03} Vadawale, S.V., Rao, A.R., Naik, S., Yadav, J.S., Ishwara-Chandra, C.H., Pramesh Rao, A. and Pooley, G.G. 2003, ApJ, 597, 1023
\end{thebibliography}
\end{document}